\documentclass[reprint,aps,prl,superscriptaddress,amsmath,longbibliography]{revtex4-2}
\usepackage{graphicx,txfonts,dcolumn,multirow,color}
\usepackage[colorlinks,linkcolor=blue,citecolor=blue,anchorcolor=blue,urlcolor=blue]{hyperref}

\begin{document}

\title{Self-similar gap dynamics in percolation and rigidity percolation}

\author{Mingzhong Lu}
\thanks{These two authors contributed equally to this work.}
\affiliation{Department of Modern Physics, University of Science and Technology of China, Hefei, Anhui 230026, China}

\author{Yu-Feng Song}
\thanks{These two authors contributed equally to this work.}
\affiliation{Department of Modern Physics, University of Science and Technology of China, Hefei, Anhui 230026, China}

\author{Ming Li}
\email{lim@hfut.edu.cn}
\affiliation{School of Physics, Hefei University of Technology, Hefei, Anhui 230009, China}

\author{Youjin Deng}
\email{yjdeng@ustc.edu.cn}
\affiliation{Department of Modern Physics, University of Science and Technology of China, Hefei, Anhui 230026, China}
\affiliation{Hefei National Research Center for Physical Sciences at the Microscale, University of Science and Technology of China, Hefei 230026, China}
\affiliation{Hefei National Laboratory, University of Science and Technology of China, Hefei, Anhui 230088, China}

\date{\today}

\begin{abstract}
Spatial self-similarity is a hallmark of critical phenomena. We investigate the dynamic process of percolation, in which bonds are incrementally inserted to an empty lattice until fully occupied, and track the gaps describing the changes in cluster sizes. Surprisingly, we find that the gap sizes follow a universal power-law distribution throughout the whole or a significant portion of process, revealing a previously unrecognized temporal self-similarity. This phenomenon appears across various percolation models, like standard, explosive and rigidity percolation. Furthermore, in rigidity percolation, we directly observe a cascading cluster-merging dynamics, triggered by single bond insertion, and further obtain a distinct temporal self-similarity in the number of merged clusters, which are hidden in static analyses. Our results also suggest that, for rigidity percolation, the temporal self-similarity is probably more intrinsic than the spatial one. These findings offer a fresh perspective on critical phenomena and broaden potential applications across complex systems.
\end{abstract}

\maketitle

Critical phenomena describe the behavior of systems undergoing continuous phase transitions~\cite{Ma2018}, where macroscopic properties change dramatically in response to small variations in control parameters. A key feature of criticality is spatial self-similarity, where structures at different scales appear statistically identical. This is characterized by the divergence of the correlation length, leading to scale invariance and universal critical exponents, which depend on dimensionality, interaction range, and symmetry.

Percolation serves as a paradigmatic model for studying critical phenomena~\cite{Stauffer1991}, traditionally focusing on configurations where bonds or sites are either present or absent. As the fraction $p$ of present bonds or sites reaches the percolation threshold $p_c$, spatial self-similarity emerges: clusters exhibit fractal geometry with a fractal dimension $d_f$ and a power-law cluster-size distribution $P_s \sim s^{-\tau}$, where $\tau$ is the Fisher exponent. Besides, scale invariance and self-similarity are also captured by finite-size scaling (FSS), with quantities like the size of the largest cluster scaling as $C_1 \sim L^{d_f}$ for systems of linear size $L$. Variations in spatial dimension or cluster formation rules, such as in explosive percolation~\cite{Achlioptas2009}, or rigidity percolation~\cite{Jacobs1995,Jacobs1996,Moukarzel1999,Briere2007}, can alter critical exponents. The scaling relation, such as $\tau=1+d/d_f$, where $d$ is the spatial dimensionality, holds across models.

Away from criticality, the power-law distribution of cluster sizes and FSS disappear, since the correlation length becomes finite. Therefore, determining critical exponents from FSS behaviors depend on the precise determination of the percolation threshold $p_c$. However, even with a high-precision $p_c$, critical behaviors can still be challenging to extract. For instance, explosive percolation, which has recently been shown to obey the standard FSS theory at a dynamic pseudocritical point~\cite{Li2023,Li2024}, was initially misidentified as a discontinuous phase transition~\cite{Friedman2009,Ziff2009,Achlioptas2009,Souza2010,Radicchi2010,Buldyrev2010} due to its abnormal critical behaviors~\cite{Grassberger2011,Souza2015}. This highlights the importance of considering the evolving nature of percolation processes, rather than relying solely on static configurations. Furthermore, in many real-world systems, percolation is often observed as a dynamic process, with applications ranging from power grids and control systems~\cite{Buldyrev2010}, to climate change~\cite{Fan2018}, traffic congestion~\cite{Li2014}, and information spreading~\cite{Xie2021}, where the dynamic aspects of percolation play a crucial role in system behaviors.

\begin{figure*}
\centering
\includegraphics[width=2.0\columnwidth]{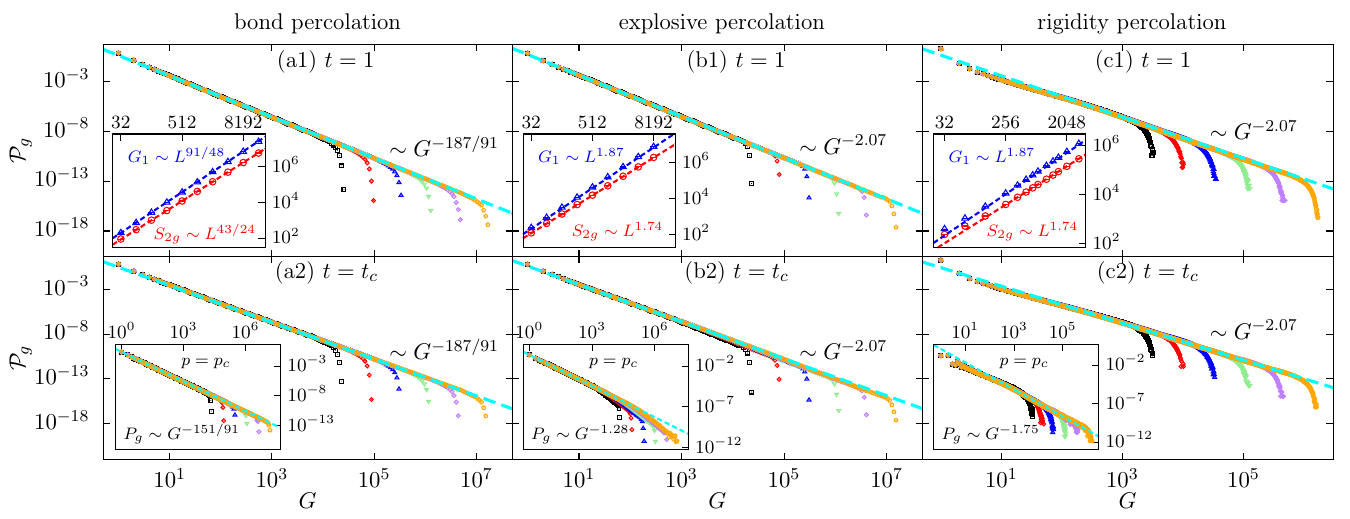}
\caption{(Color online) Dynamic self-similarity in the gap distribution $\mathcal{P}_g$ is demonstrated for (a) bond percolation on square lattices, (b) explosive percolation on complete graphs, and (c) rigidity percolation on triangular lattices. In (a1)-(c1), gaps are collected from $t=0$ to $t=1$, while in (a2)-(c2), gaps are collected up to $t=t_c<1$. For all these cases, a nice power-law distribution, $\mathcal{P}_g\sim G^{-\tau}$, where $\tau$ is the Fisher exponent, is observed. The insets of (a1)-(c1) show the largest gap $G_1\sim L^{d_f}$, and the susceptibility-like quantity $S_{2g}\sim L^{2d_f-d}$. The insets of (a2)-(c2) show the gap distribution $P_g$ sampled at $p_c$. A similar power-law, $P_g\sim G^{-\tau'}$, is observed, where $\tau'=1+X_2/d_f<\tau$. For (a) and (b), scatters of different colors (shapes) refer to systems of different side lengths from $L=2^8$ to $2^{13}$, while for (c) are those from $L=2^6$ to $2^{11}$. For complete graphs, the side length is defined as $L\equiv \sqrt{V}$, and the rescaled time step as $t\equiv T/V$, with $V$ the number of sites. (a) $t_c=p_c=1/2$, $d_f=91/48$, $X_2=5/4$, $\tau=187/91$, and $\tau'=1+X_2/d_f=151/91$. (b) The fit results in Refs.~\cite{Li2023,Li2024} give $t_c=p_c=0.8884491$, $d_f=1.87$ and $\tau=2.07$. (c) Our fit gives consistent values $d_f=1.870(8)$ and $2d_f-d=1.738(8)$. The Fisher exponent is given by $\tau=1+d/d_f=2.07$. The percolation threshold is $t_c=p_c=0.660254(3)$.} \label{f1}
\end{figure*}

In this Letter, we reveal a surprising phenomenon: by interpreting percolation as gap increment dynamics, we uncover a universal temporal self-similarity throughout the whole or a significant portion of process, without knowing the exact value of $p_c$. Specifically, for bond percolation, this gap dynamics is realized by randomly inserting bonds one by one, causing the growth of clusters. At a rescaled time step $t\equiv T/E$, where $T$ denotes the time step (number of inserted bonds) and $E$ is for the total number of bonds, we record the increment in cluster size, denoted as a gap $\mathcal{G}(t)$. As shown in Fig.~\ref{f1} (a1) for square lattices of side length $L$, the gap distribution $\mathcal{P}_g$, collected from the whole process (from $t=0$ to $t=1$), follows a power-law, $\mathcal{P}_g \sim G^{-\tau}$, where $\tau=187/91$ is the Fisher exponent in two dimensions (2D). Additionally, observables like the largest gap $G_1\equiv \langle \mathcal{G}_1\rangle$, where $\langle\cdot\rangle$ represents the average over realizations, and the second moment of the gap distribution $S_{2g}\equiv \langle\sum_{t=0}^1 \mathcal{G}(t)^2\rangle/L^d$ scale similarly to their counterparts in cluster-size distribution of a static configuration, i.e., $G_1 \sim L^{d_f}$ and $S_{2g}\sim L^{2d_f-d}$. Remarkably, this dynamic self-similarity persists, as long as the gap dynamics does not terminate before $t_c=p_c=1/2$. Moreover, this dynamic self-similarity can be observed for site percolation (End Matter). For explosive percolation, a clean FSS can also be revealed without abnormal critical behaviors (Fig.~\ref{f1} (b)).

Particularly intriguing is rigidity percolation, where a mechanically stable spanning cluster is required to transmit stress across the system. Although the rigid cluster is widely used to describe connectivity of various systems, such as molecular glasses~\cite{Thorpe2000}, gels~\cite{Broedersz2011,Zhang2019}, jamming transition~\cite{Henkes2016}, granular systems~\cite{Dashti2023}, fibers~\cite{Vinutha2023}, and living tissues~\cite{Petridou2021,Rozman2024}, the critical behavior of rigidity percolation remains significantly less well understood. The determination of precise critical exponents and the identification of universality classes have been topics of ongoing debate~\cite{Feng1984,Kantor1984,Hansen1989,Jacobs1995,Jacobs1996,Moukarzel1999,Plischke2007,Briere2007,Zhang2015,Ellenbroek2015}. Using gap dynamics, we reveal that, in addition to exhibiting a well-defined gap scaling with self-consistent $d_f$ and $\tau$, rigidity percolation demonstrates a cascade cluster-merging process triggered by a single bond insertion, where the number of merged clusters $\mathcal{K}$ can be very large. Similar to gap distribution, $\mathcal{K}$ also has dynamic self-similarity, demonstrating a power-law distribution, $\mathcal{P}_K\sim K^{-\tau_K}$. Besides, while the mean value $K(t)\equiv\langle\mathcal{K}(t)\rangle$ for any given $t$ shows a finite value, the maximum number of merged clusters during gap dynamics diverges as a power-law $K_{\text{max}}\sim L^{d_K}$, with hyperscaling relation $\tau_K=1+d/d_K$.

In contrast to the clean dynamic gap self-similarity, the static cluster-size distribution $P_s$ in rigidity percolation shows a Fisher exponent $\tau<2$ for linear sizes up to $L=2048$. This violation of normalization condition, as well as the hyperscaling relation, may stem from complex finite-size corrections. Additionally, the values of $d_f$ determined from dynamic and static analyses are not fully consistent within their error bars. We argue that, due to the cascade effect, the dynamic self-similarity is more robust than the static scaling, the $d_f$ value determined from gap dynamics is likely more reliable. We also conjecture that $d_K=1/\nu$, with $\nu$ the correlation-length exponent. Recent studies have provided evidence for conformal invariance in rigidity percolation \cite{Javerzat2023}, indicating a connection to Schramm-Loewner evolution theory~\cite{Javerzat2024}. It remains an open question how to incorporate the cascade effect within these field-theoretical treatments.

\emph{Self-similar Gap Dynamics} -- In general, percolation models can be interpreted as dynamics of cluster growth, where bonds or sites are randomly inserted. At each time step $T$, the insertion of a bond or site may cause the merging of $\mathcal{K}$ clusters, resulting a large cluster of size $\mathcal{C}(T+1)$. For bond percolation, $\mathcal{K}$ can only be $1$ or $2$, while in site percolation or other extended models, $\mathcal{K}$ can be larger than $2$. We define a gap for cluster growth as
\begin{equation}
\mathcal{G}(t) \equiv \mathcal{C}(T+1) -\max(\mathcal{C}_1(T),\mathcal{C}_2(T),\ldots,\mathcal{C}_\mathcal{K}(T)),     \label{eq-g}
\end{equation}
where $\mathcal{C}_i(T)~(i=1,2,\ldots,\mathcal{K})$ are the sizes of the clusters to be merged at $T$, and $\max(\cdot)$ denotes the largest one among these clusters. For $t<t_c$, small clusters lead to low values of $\mathcal{G}(t)$. As $t$ increases, larger clusters form, causing $\mathcal{G}(t)$ to grow. Above $t_c$, the formation of a giant cluster leaves smaller clusters behind, resulting in a decrease in $\mathcal{G}(t)$.

We also consider two other important and challenging percolation models: explosive and rigidity percolation. For explosive percolation, at each time step, two potential bonds are chosen randomly, and the bond that minimizes the product of the sizes of the associated clusters is inserted~\cite{Achlioptas2009,Li2023}. Explosive percolation exhibits rich critical phenomena, attracting significant research attention~\cite{Achlioptas2009,Friedman2009,Ziff2009,Radicchi2010,Souza2010,Grassberger2011,Souza2015,Yang2024}. Recent studies have shown that it continues to obey the standard finite-size scaling theory when an event-based ensemble is applied~\cite{Li2023,Li2024}. Rigidity percolation follows the same bond-insertion rule as bond percolation but focuses exclusively on the formation of rigid clusters capable of transmitting stress~\cite{Jacobs1995,Jacobs1996,Moukarzel1999,Briere2007}. However, the merged-cluster number $\mathcal{K}$ could be much larger than $2$. By interpreting the formation of rigid clusters as a pebble game~\cite{Jacobs1995,Jacobs1997}, we develop an algorithm for dynamically tracking cluster growths~\cite{Lu2024}, with small additional computational overhead on the identification of rigid clusters. The simulation is performed up to $L=3072$ on the triangular lattice.

In Figs.~\ref{f1} (a1)-(c1), we present the distribution $\mathcal{P}_g$ of gaps collected from $t=0$ to $t=1$ for bond percolation on square lattices, explosive percolation on complete graphs, and rigidity percolation on triangular lattices. For complete graphs, we artificially define a side length $L\equiv\sqrt{V}$, where $V$ is the total number of sites, and the rescaled time step is defined as $t\equiv T/V$. All these systems exhibit a well-defined scaling $\mathcal{P}_g \sim G^{-\tau}$, for which the self-similarity is temporal and does not require a priori knowledge of $p_c$. This observation does not require the entire process to evolve, and is not sensitive to the time where the dynamic process is terminated, as long as $t_c$ is reached. This is demonstrated in Figs.~\ref{f1} (a2)-(c2), when gaps are collected only up to $t=t_c<1$. This highlights a universal dynamic self-similarity that persists beyond the static configuration.

Key observables -- the largest gap, $G_1$, and a susceptibility-like quantity, $S_{2g}$ -- also exhibit interesting FSS. In the insets of Figs.~\ref{f1} (a1)-(c1), both observables display clear FSS. For 2D bond percolation ($d_f=91/48$), the observations suggest the finite-size scaling relations: $G_1 \sim L^{d_f}$ and $S_{2g} \sim L^{2d_f-d}$, consistent with the scaling $\mathcal{P}_g \sim G^{-\tau}$ under hyperscaling relation $\tau=1+d/d_f$. For explosive percolation, where $d_f=1.87$~\cite{Li2023,Li2024}, this scaling relation are similarly consistent, without any abnormal FSS behaviors as seen in static analyses~\cite{Grassberger2011,Souza2015}. For rigidity percolation, the fit to the FSS ansatz
\begin{equation}
Q(L)=L^{Y_Q}(a_0+a_1L^{-\omega_1}+a_2L^{-\omega_2}),     \label{eq-sa}
\end{equation}
with finite-size corrections $L^{-\omega_i} (i=1, 2)$, yield consistent values $d_f=1.870(8)$ for $G_1$ and $2d_f-d=1.738(8)$ for $S_{2g}$.

\begin{figure}
\centering
\includegraphics[width=1.0\columnwidth]{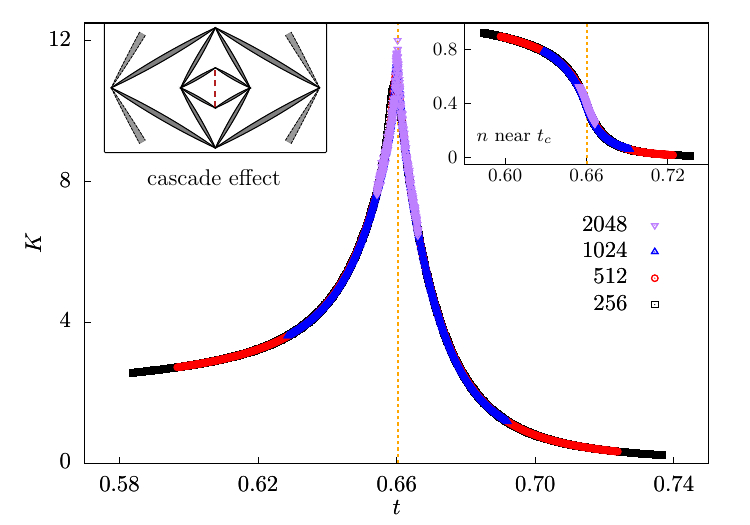}
\caption{(Color online) The mean number of merged clusters, $K(t)\equiv\langle\mathcal{K}(t)\rangle$, as a function of time step $t$, for rigidity percolation on triangular lattices of different side lengths $L$. The curves of different $L$ nearly collapse on a single curve, with peak at $t_c=0.660254(3)$ (indicated by the dashed line). The left inset illustrates a cascade process triggered by the insertion of a single bond. Gray lines represent independent rigid clusters, and when a new bond is inserted (depicted by the red dashed line), nonrigid connections between these clusters become rigid one after another, propagating from the interior to the exterior. The right inset displays the decreasing of the number of clusters per site $n$ as $t$ increasing, showing hardly any finite-size behaviors.} \label{f2}
\end{figure}

We also sample gaps for a static ensemble at $p_c$. To distinguish from the dynamic distribution $\mathcal{P}_g$, we denote this static gap distribution as $P_g$. As seen in the insets of Figs.~\ref{f1} (a2)-(c2), $P_g$ exhibits a power-law form $P_g \sim G^{-\tau'}$, with $\tau'<\tau$, being a new Fisher exponent. Using the $2$-cluster exponent $X_2$~\cite{Saleur1987}, also known as $4$-path crossing exponent~\cite{Aizenman1999} or $4$-arm exponent~\cite{Smirnov2001}, we derive $\tau'=1+X_2/d_f$ (End Matter).

In 2D percolation, with $X_2=5/4$ and $d_f=91/48$, this yields $\tau'=151/91$, which aligns with simulation results in the inset of Fig.~\ref{f1} (a2). This exponent $\tau'$ and the scaling relation $\tau'=1+X_2/d_f$ hold universally across site and bond percolation models, as shown in End Matter. For explosive and rigidity percolation, where $X_2$ remains unknown, simulations (Fig.~\ref{f1} (b2) and (c2)) suggest values of $\tau'=1.28$ and $1.75$, respectively. These values allow us to estimate $X_2$ for these models, giving $X_2=1.4$ for rigidity percolation in 2D and $X_2=0.26$ in terms of volume for explosive percolation on complete graphs.

\begin{figure}
\centering
\includegraphics[width=1.0\columnwidth]{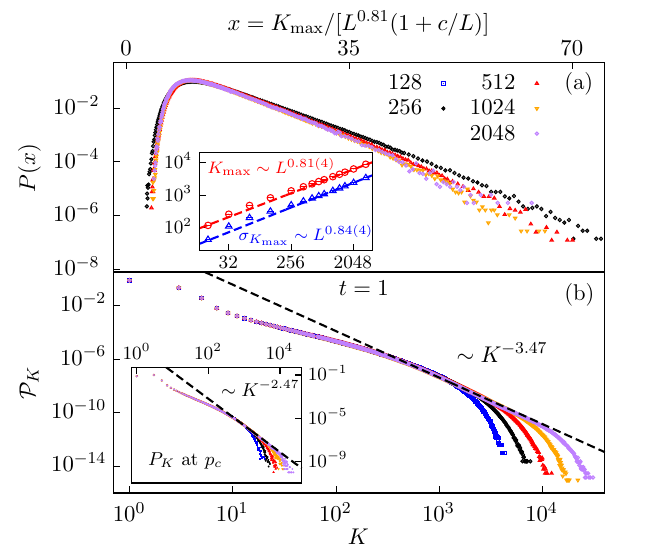}
\caption{(Color online) Cascade dynamics in rigidity percolation and its self-similar scaling in the distribution of merged-clusters number. (a) The distribution $P(x)$ of the maximum number $\mathcal{K}_{\text{max}}$ of merged clusters during gap dynamics, where $x\equiv K_{\text{max}}/[L^{0.81}(1+cL^{-1})]$ with $c=158$. The inset shows the FSS of $K_{\text{max}}$ and its fluctuation $\sigma_{K_{\text{max}}}$, with the lines indicating the fit results. (b) The distribution $\mathcal{P}_K$ of $\mathcal{K}$ during gap dynamics. A power-law distribution $\mathcal{P}_K\sim K^{-\tau_K}$ with $\tau_K=1+d/d_K=3.47$. The inset provides the distribution $P_K$ at $p_c$, following a power-law $P_K\sim K^{-\tau_K+1}$.} \label{f3}
\end{figure}

\emph{Self-similar Cascade Dynamics in Rigidity Percolation} -- In the conventional static study of rigidity percolation, all occupied bonds are randomly placed on a lattice first, and then geometric clusters are constructed according to the rigidity rules using the pebble game algorithm~\cite{Jacobs1995,Jacobs1997}. In contrast, gap dynamics reveals a dynamic feature: the large number of merged clusters, $\mathcal{K}$, arising from cascade processes during rigid cluster formation. As shown in the left inset of Fig.~\ref{f2}, when a new bond (dashed red line) is added, certain nonrigid connections between rigid clusters (solid gray lines) become rigid, transforming an entire nonrigid subgraph into a rigid structure. This transformation can trigger additional nonrigid connections to solidify, creating a cascade effect.

This cascade effect cannot be seen in the static analyses. As in the right inset of Fig.~\ref{f2}, the mean cluster number per site $n$ is a smooth function of $t$, independent of $L$. In gap dynamics, one can track the number of merged clusters $\mathcal{K}(t)$ as a function of $t$. It is shown in Fig.~\ref{f2} that $K(t)\equiv \langle\mathcal{K}(t)\rangle$ has a point peak at $t_c$. However, $K(t)$ does not seem to have nontrivial FSS behaviors -- $K(t)$ of different system sizes collapse on a single curve with a nondivergent $K(t_c)$.

Nevertheless, the behavior of $K(t)$ near $t_c$ allows us to identify the time step at which $\mathcal{K}$ reaches its maximum $\mathcal{K}_{\text{max}}$, as a pseudocritical point $p_\mathcal{K}$. Then, the critical point $p_c=t_c=0.660254(3)$ is obtained by fitting the mean $\langle p_\mathcal{K}\rangle$ to the FSS ansatz, $\langle p_\mathcal{K}\rangle=p_c+L^{-y_0}(a_0+a_1L^{-y_1})$, with $a_0$ and $a_1$ as $L$-independent parameters.

Along with the identification of $p_\mathcal{K}$, we observe the divergence of $K_{\text{max}}\equiv\langle\mathcal{K}_{\text{max}}\rangle$ as system size increases, distinguishing itself from behaviors near $t_c$. In the inset of Fig.~\ref{f3} (a), both $K_{\text{max}}$ and its fluctuation $\sigma_{K_{\text{max}}}\equiv
\sqrt{\langle\mathcal{K}_{\text{max}}^2\rangle-\langle\mathcal{K}_{\text{max}}\rangle^2}$ exhibit the FSS behavior, $\sim L^{d_K}$. The fit to the FSS ansatz Eq.~(\ref{eq-sa}) yields a consistent value $d_K=0.81(4)$. Defining $x\equiv K_{\text{max}}/[L^{d_K}(1+cL^{-1})]$, where the term $cL^{-1}$ with $c=158$ accounts for the strong finite-size correction, the distribution $P(x)$ trends to collapse onto a single curve for large systems (Fig.~\ref{f3} (a)), confirming the well-defined scaling of $K_{\text{max}}$.

As for gap sizes, the distribution of $\mathcal{K}$ during gap dynamics also exhibits a scaling behavior $\mathcal{P}_K\sim K^{-\tau_K}$ (Fig.~\ref{f3} (b)). For large systems, simulation results suggest $\tau_K=3.47$, satisfying the hyperscaling relation $\tau_K=1+d/d_K$ for $d_K=0.81$. As shown in the inset of Fig.~\ref{f3} (b), the static distribution $P_K$ of $\mathcal{K}$ at $p_c$ also follows a power-law distribution, $P_K\sim K^{-\tau_K+1}$. From this, the mean merged-cluster number at $p_c$ can be estimated as $K_c \sim \int_{K=1}^{L^{d_K}} K^{2-\tau_K} dK \sim aL^{(3-\tau_K)d_K} +\mathcal{O}(1)$, where $a$ is an $L$-independent coefficient. Since $\tau_K>3$, it has $K_c\sim \mathcal{O}(1)$, explaining the behaviors in Fig.~\ref{f2}.

\begin{figure}
\centering
\includegraphics[width=1.0\columnwidth]{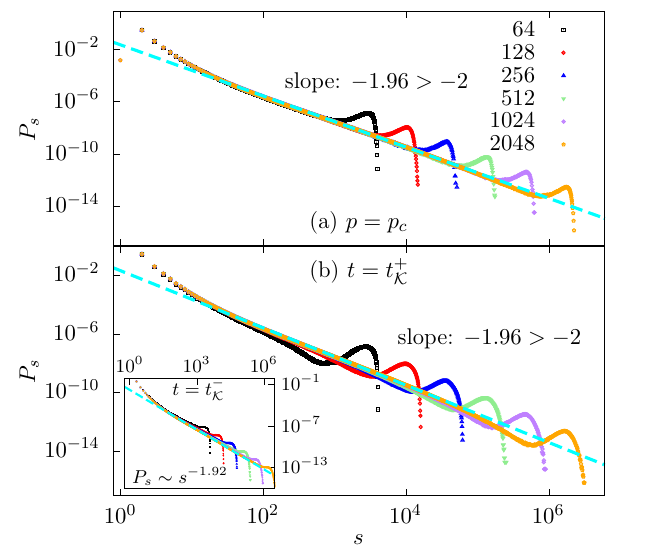}
\caption{(Color online) Unsatisfactory spatial self-similarity of cluster-size distribution $P_s$ in rigidity percolation for systems of different side lengths $L$. (a) At the critical point $p_c$. (b) At the dynamic pseudocritical point $t_\mathcal{K}^+$, after merging $\mathcal{K}_{\text{max}}$ clusters. The inset of (b) shows $t_\mathcal{K}^-$, before merging $\mathcal{K}_{\text{max}}$ clusters. In all cases, the power-law scaling $P_s\sim s^{-\tau}$ is observed, with $\tau<2$.} \label{f4}
\end{figure}

\emph{Unsatisfactory Spatial Self-similarity in Rigidity Percolation} -- While dynamic self-similarity is observed in gap dynamics (Fig.\ref{f1} (c)), the static configuration at $p_c$ also exhibits spatial self-similarity, as the cluster-size distribution follows $P_s\sim s^{-\tau}$ (Fig.\ref{f4} (a)). However, with a Fisher exponent $\tau<2$, the normalization condition is violated, and the hyperscaling relation $\tau=1+d/d_f$ is broken.

We further investigate $P_s$ at the dynamic pseudocritical point $p_\mathcal{K}$, where fluctuations among different realizations are minimized. Regardless of whether the configuration is taken just before or after the merging of $\mathcal{K}_{\text{max}}$ clusters (denoted as $t_\mathcal{K}^-$ and $t_\mathcal{K}^+$), spatial self-similarity remains unsatisfactory, with $P_s$ continuing to yield $\tau<2$ (Fig.~\ref{f4} (b) and its inset). Moreover, although the bond configurations at $t_\mathcal{K}^-$ and $t_\mathcal{K}^+$ differ only by one bond, $P_s$ differs significantly, emphasizing the fragility of spatial self-similarity, and the crucial role of the cascade effect in rigidity percolation. Based on these observations, we argue that the dynamic gap self-similarity is probably more robust and intrinsic than the spatial one. The violation of the normalization and the hyperscaling relation in static analyses might be cured for sufficiently large systems, which is unfortunately beyond our current capability.

\emph{Discussion} -- We have demonstrated a universal dynamic self-similarity across percolation systems through gap dynamics, revealing critical behaviors as a gap-increment process. In contrast to static analyses which focuses on the spatial property at criticality, the dynamic self-similarity is a cumulative effect over the whole critical region. Thus, it is insensitive to the termination of the percolation process as long as criticality is reached. In explosive percolation, the self-similar gap dynamics is very clean, where the previous observed anomalous critical phenomena in static analyses are completely absent, allowing a precise determination of fractal dimension and universality.

The dynamic analyses allows to reveal critical behaviors that cannot or are difficult to be observed in static analyses. In rigidity percolation, the self-similar dynamics in both the gap size and the number of merged clusters are observed. We note that at different pseudocritical points, the values of the fractal dimension from static analyses are not fully consistent with each other~\cite{Jacobs1995,Moukarzel1999,Lu2024}, and they also differ somewhat from the dynamic estimation. Taking into account the complication in spatial self-similarity due to cascade effect, we argue that the dynamic estimate, $d_f=1.870(8)$, is more reliable. Finally, we also conjecture $d_K=1/\nu$, with $d_K=0.81(4)$ the FSS exponent for the maximum number of merged clusters and $\nu$ the correlation-length exponent, consistent with the existing results $\nu=1.21(6)$~\cite{Jacobs1995}, and $\nu=1.19(1)$~\cite{Javerzat2023}.

Gap dynamics reinterprets core insights of percolation theory, offering a dynamic approach to observing critical phenomena that overcomes limitations of static analysis across varied scenarios. Beyond percolation systems, this framework provides an adaptable and precise method for identifying critical behaviors, opening doors to applications in real-world systems with inherent time-evolving behaviors, from infrastructure networks to biological and social systems.

We are aware that a recent work~\cite{Fang2024} independently addresses a similar question but from a significantly different perspective. The research was supported by the National Natural Science Foundation of China under Grant No.~12275263, the Innovation Program for Quantum Science and Technology under Grant No.~2021ZD0301900, and Natural Science Foundation of Fujian province of China under Grant No.~2023J02032.

\bibliography{ref}

\appendix

\section{End Matter}

\begin{figure}
\centering
\includegraphics[width=1.0\columnwidth]{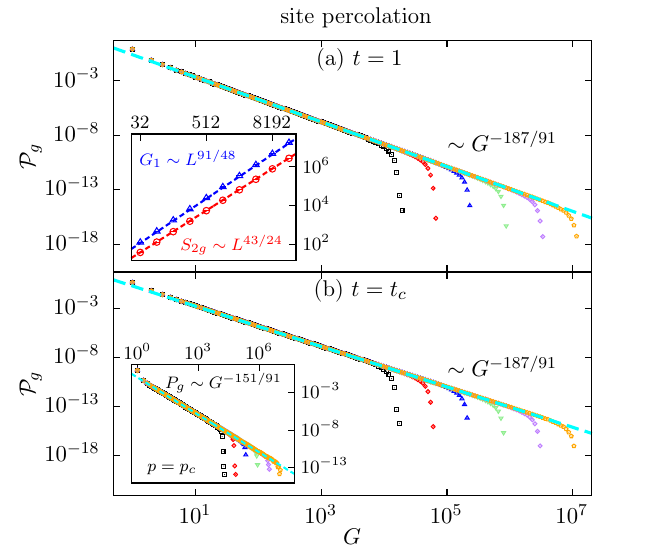}
\caption{(Color online) Dynamic self-similarity in the gap distribution $\mathcal{P}_g$ for site percolation on square lattices. Scatter points in various colors and shapes represent systems with side lengths ranging from $L=2^8$ to $L=2^{13}$. Here, the rescaled time step is defined as $t\equiv T/E$. In (a), gaps are collected from $t=0$ to $t=1$, while in (b), only up to $t_c=p_c=0.59274605\ldots$~\cite{Jacobsen2015}. In both cases, a power-law $\mathcal{P}_g \sim G^{-\tau}$ with $\tau=187/91$ is observed. Inset of (a) shows the largest gap $G_1\sim L^{d_f}$ and susceptibility-like quantity $S_{2g}\sim L^{2d_f-d}$, with $d_f=91/48$. Inset of (b) shows the gap distribution $P_g\sim G^{-\tau'}$ at $p_c$, with $X_2=5/4$ and $\tau'=1+X_2/d_f=151/91$.} \label{f5}
\end{figure}

\begin{figure}
\centering
\includegraphics[width=1.0\columnwidth]{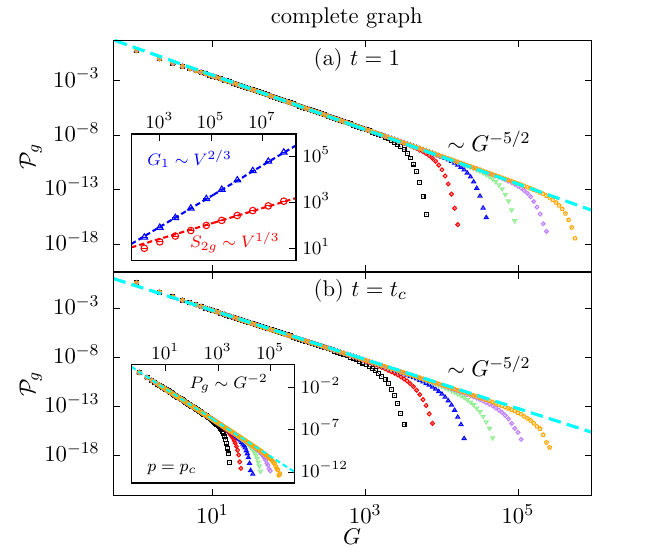}
\caption{(Color online) Dynamic self-similarity in the gap distribution $\mathcal{P}_g$ for bond percolation on complete graphs. Scatter points in various colors and shapes represent systems with volumes ranging from $V=2^{16}$ to $V=2^{26}$. Here, the rescaled time step is defined as $t\equiv T/V$. In (a), gaps are collected from $t=0$ to $t=1$, while in (b), only up to $t_c=p_c=0.5$. In both cases, a power-law $\mathcal{P}_g \sim G^{-\tau}$ with $\tau=5/2$ is observed. Inset of (a) shows the largest gap $G_1\sim V^{d_f}$ and susceptibility-like quantity $S_{2g}\sim V^{2d_f-1}$, with $d_f=2/3$. Inset of (b) presents the gap distribution $P_g\sim G^{-\tau'}$ at $p_c$, with $X_2=2/3$ and $\tau'=1+X_2/d_f=2$.} \label{f6}
\end{figure}

\begin{figure}
\centering
\includegraphics[width=1.0\columnwidth]{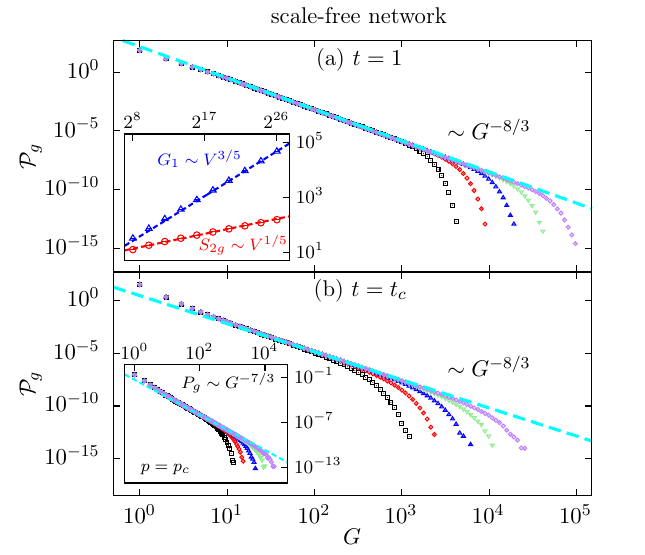}
\caption{(Color online) Dynamic self-similarity in the gap distribution $\mathcal{P}_g$ for bond percolation on scale-free networks. Scatter points in various colors and shapes represent systems with volumes ranging from $V=2^{16}$ to $V=2^{24}$. Here, the rescaled time step is defined as $t\equiv T/E$.  In (a), gaps are collected from $t=0$ to $t=1$, while in (b), only up to the critical point, $t_c=p_c=\langle k\rangle/\langle k(k-1)\rangle$~\cite{Molloy1995,Callaway2000}, where $k$ denotes node degree, and $\langle \cdot\rangle$ denotes an average over the network. Here, we consider networks with minimum degree $k_{\text{min}}=2$ and $\lambda=3.5$, thus, the critical point is $t_c=p_c=0.2686997\ldots$. Both cases reveal a power-law scaling, $\mathcal{P}g \sim G^{-\tau}$, with $\tau=8/3$. The inset of (a) shows finite-size scaling for the largest gap, $G_1\sim V^{d_f}$, and a susceptibility-like quantity, $S_{2g}\sim V^{2d_f-1}$, where $d_f=3/5$. The inset of (b) presents the gap distribution at $p_c$, $P_g\sim G^{-\tau'}$, with $X_2=4/5$ and $\tau'=1+X_2/d_f=7/3$.} \label{f7}
\end{figure}

\emph{Gap Dynamics in Other Percolation Models} -- Beyond the percolation models shown in Fig.~\ref{f1}, we analyze gap dynamics in site percolation on square lattices (Fig.~\ref{f5}), bond percolation on complete graphs (Fig.~\ref{f6}), and bond percolation on scale-free networks (Fig.~\ref{f7}). In each case, the time step is defined as $t\equiv T/V$, where $T$ is the number of inserted sites or bonds, and $V$ is the total number of sites (system volume). The results demonstrate that the universal scaling behaviors identified through gap dynamics in Fig.~\ref{f1} hold consistently across these models.

For site percolation in 2D (Fig.\ref{f5}), we observe scaling behaviors with critical exponents that match those of bond percolation in 2D (Fig.\ref{f1}). In bond percolation on complete graphs, where the finite-size scaling is in terms of $V$, we argue that the $2$-cluster exponent $X_2$ is double that of the $1$-cluster exponent $X_1=1-d_f$, reflecting that two clusters branching from the two ends of a bond are effectively independent. With $d_f=2/3$, we obtain $X_2=2(1-d_f)=2/3$ and thus $\tau'=1+X_2/d_f=2$, while the Fisher exponent for bond percolation on complete graphs is $\tau=5/2$, all of which align with the scaling in gap dynamics (Fig.~\ref{f6}).

For scale-free networks with degree distribution $\sim k^{-\lambda}$ (where $k$ is node degree), we examine the case of $\lambda=3.5$ in Fig.\ref{f7}. In bond percolation on such networks with $\lambda>3$, the critical exponents are $\tau=(2\lambda-3)/(\lambda-2)$ and $d_f=(\lambda-2)/(\lambda-1)$~\cite{Cohen2002}, where the finite-size scaling is also in terms of $V$. For $\lambda=3.5$, this yields $\tau=8/3$ and $d_f=3/5$, both of which are consistent with our observations. Similar to complete graphs, we argue that the relation $X_2=2X_1=2(1-d_f)$ also holds for scale-free networks, giving $X_2=2/(\lambda-1)$. For $\lambda=3.5$, this results in $X_2=4/5$ and thus $\tau'=7/3$, in agreement with simulation data shown in the inset of Fig.~\ref{f7}(b).

\emph{Derivation of the Scaling Relation $\tau'=1+X_2/d_f$} -- From $P_g \sim G^{-\tau'}$, with the observation that $\tau'<2$, the mean gap at $p_c$ can be written as
\begin{equation}
\langle G \rangle \sim \int_1^{L^{d_f}} G^{1-\tau'} dG \sim L^{d_f(2-\tau')}. \label{eq-mg}
\end{equation}
By the definition of gap in Eq.~(\ref{eq-g}), when two clusters merge due to a newly inserted bond, the gap just represents the size of the smaller cluster, which scales as $\sim L^{d_f-X_2}$~\cite{Deng2010}. Comparing with Eq.~(\ref{eq-mg}), we derive the scaling relation
\begin{equation}
\tau' = 1 + \frac{X_2}{d_f}. \label{eq-tau}
\end{equation}
This relation does not depend on any specific percolation model, indicating that Eq.~(\ref{eq-tau}) represents a universal scaling relation, valid across different percolation models.

\end{document}